\documentclass[twocolumn,draft,prd,nofootinbib]{revtex4}
\usepackage{epsf}
\usepackage{dcolumn}

\newcommand{\be}{\begin{equation}}
\newcommand{\ee}{\end{equation}}
\newcommand{\Deln}{\ensuremath{\Delta N_\nu\;}}
\newcommand{\epm}{\ensuremath{e^{\pm}\;}}

\def\ie{{\it i.e.},~}
\def\eg{{\it e.g.},~}
\def\etal{{\it et al.}~}
\def\4he{$^4$He}
\def\3he{$^3$He}
\def\7li{$^7$Li}
\def\Yp{Y$_{\rm P}$~}
\def\yd{$y_{\rm D}$~}
\def\yli{$y_{\rm Li}$~}
\def\xie{$\xi_{e}$~}
\def\hii{H\thinspace{$\scriptstyle{\rm II}$}~}
\def\Nnu{N$_{\nu}$~}

\newcommand\la{\lower0.6ex\vbox{\hbox{\ensuremath{\buildrel{\textstyle<}\over{\sim}\ }}}}
\newcommand\ga{\lower0.6ex\vbox{\hbox{\ensuremath{\buildrel{\textstyle>}\over{\sim}\ }}}}

\newcommand{\obh}{\ensuremath{\Omega_{\rm B} h^2\;}}

\begin{document}

\title{BBN For Pedestrians}

\author{James P. Kneller$^1$ and Gary Steigman$^{2,3}$\\}

\affiliation{$^1$Department of Physics, North Carolina State
             University, Raleigh, NC 27695-8202}
             
\affiliation{$^2$Department of Physics, The Ohio State University, Columbus, 
             OH 43210~\footnote{mailing address}}
             
\affiliation{$^3$Department of Astronomy, The Ohio State University, Columbus, 
             OH 43210}

\date{{\today}\\}


\begin{abstract}

The simplest, ``standard'' model of Big Bang Nucleosynthesis (SBBN) assumes 
three light neutrinos (\Nnu = 3) and no significant electron neutrino 
asymmetry ($\nu_{e}-\bar{\nu}_{e}$ asymmetry parameter $\xi_{e} \equiv 
\mu_{e}/kT$, where $\mu_{e}$ is the $\nu_{e}$ chemical potential) leaving 
only one adjustable parameter: the baryon to photon ratio $\eta \equiv 
n_{\rm B}/n_{\gamma}$.  The primordial abundance of any one nuclide can, 
therefore, be used to measure $\eta$ and the value derived from the 
observationally inferred primordial abundance of deuterium closely matches 
that from current non-BBN data, primarily from the WMAP survey.  However, 
using this same estimate, there is a tension between the SBBN-predicted 
helium-4 and lithium-7 abundances and their current, observationally 
inferred primordial abundances, suggesting that \Nnu may differ from the 
standard model value of three and/or that $\xi_{e}$ may differ from zero 
(or, that systematic errors in the abundance determinations have been 
underestimated or overlooked).  The differences are not large and the 
allowed ranges of the BBN parameters ($\eta$, $N_{\nu}$, and $\xi_{e}$) 
permitted by the data are quite small. Within these ranges, the 
BBN-predicted abundances of D, \3he, \4he, and \7li are very smooth, 
monotonic functions of $\eta_{10}$, \Deln $ \equiv$ N$_{\nu}-3$, and 
$\xi_{e}$.  As a result, it is possible to describe the dependencies 
of these abundances (or powers of them) upon the three parameters by 
simple, {\bf{\it linear}} fits which, over their ranges of applicability, 
are accurate to a few percent or even better.  The fits presented here 
have {\it not} been maximized for their accuracy but, rather, for their 
{\it simplicity}.  To identify the ranges of applicability and relative 
accuracies, they are compared to detailed BBN calculations; their utility 
is illustrated with several examples.  Given the tension within BBN, 
these fits should prove useful in facilitating studies of the viability 
of various options for non-standard physics and cosmology, prior to 
undertaking detailed BBN calculations.  

\end{abstract}

\pacs{}
\keywords{Suggested keywords}

\maketitle


\section{~Big Bang Nucleosynthesis}

Shortly after the emergence of the Big Bang model it was realized that 
conditions were ripe during the early Universe for a brief period of 
nucleosynthesis. Just as in every other setting where nuclear reactions 
occur, the yields of the emerging nuclides are governed by three 
environmental characteristics: the duration of the event, the density 
of the reactants, and their thermal properties.  A brief, dilute and 
cool environment would yield a very different set of abundances compared 
to those that would emerge if Big Bang Nucleosynthesis (BBN) had been 
long-lasting, dense and hot.  Although it is possible to characterize 
BBN in such broad generality (with some success; see, for example,
\cite{CK2002,MR2003}), the paradigm most frequently encountered 
uses the Friedman equation to relate the BBN expansion rate, $H$, 
to the thermal properties of the particles present at that epoch
\begin{equation}
H^{2} = \frac{8\pi G_{N}}{3}\;\rho,
\end{equation}
with $G_{N}$ being Newton's constant and $\rho$ the total energy density. 
The standard model of particle physics provides the candidate particles 
whose energy density contribute to $\rho$, but it is possible that this 
may fall short of the actual particle content at that time.  Even so, 
after blueshifting the currently observed Cosmic Background Radiation 
(CBR) energy density back to the epoch of BBN, it emerges that the energy 
density of the Universe was dominated during BBN by the relativistic 
particles.  The thermal properties of each particle species are described 
by a temperature and a chemical potential, but thermal coupling among 
the constituents equilibrates the temperatures while chemical coupling 
relates the chemical potentials.  Charge neutrality provides additional 
relations.  This reduces the number of independent quantities with the 
result that BBN can be described by a minimal set of parameters, typically 
taken to be: the density of the nucleons/baryons, the chemical potentials 
of the three, active neutrinos, and the energy density in unaccounted 
particles (and/or additional terms on the right hand side of eq.~1).  
In what follows the role of each is discussed and the parameters used 
to describe them are introduced.

\subsection{~The Baryon Density: $\eta$}

The simplest, ``standard" model for BBN (SBBN) sets any ``extra" energy 
density and the chemical potentials of the three neutrinos to zero leaving 
just the density of the baryons, $n_{\rm B}$, as the only adjustable/free 
parameter.  But since $n_{\rm B}$ drops as the Universe expands, it is useful 
to introduce the baryon to photon ratio $\eta \equiv n_{\rm B}/n_{\gamma}$, 
($\eta_{10} \equiv 10^{10}\eta$), since this quantity is dimensionless and
constant, except during the period of \epm annihilation.  

From inspection of the flow of nuclei through the reaction network in SBBN 
it can be seen that the primordial abundances of the relic light nuclides 
D, \3he, and \7li are determined by the competition between the nuclear 
production and destruction rates, which scale with the nucleon density. 
To an extent which varies from nuclide to nuclide, these three nuclei 
are all potential baryometers.  In contrast, as the light nucleus with 
the largest binding energy per nucleon, the \4he relic abundance is 
relatively {\it insensitive} to the magnitudes of the nuclear reaction 
rates because, for the range in baryon density of interest here, they are 
sufficiently rapid to burn nearly all neutrons present initially to \4he. 
What sensitivity it does possess is due to the role that $\eta$ plays in 
determining the temperature at which the D abundance becomes significant: 
it is the build up of a large D abundance that leads to the most rapid 
phase of nuclear burning in BBN and the termination of free neutron decay 
as an important process. 


\subsection{Early-Universe Expansion Rate: $S$}

The early Universe is radiation-dominated, with its expansion rate determined
by the energy density in the relativistic particles, $\rho_{\rm R}$.  Prior 
to BBN, \eg at a temperature of a few MeV, before \epm annihilation, the 
standard model of particle physics provides photons, \epm pairs and three 
flavors of left-handed (\ie one helicity state) neutrinos (and their 
right-handed, antineutrinos) as constituents of this dominant component. 
With all chemical potentials set to zero the energy densities are related 
by thermal equilibrium so that  
\be
\rho_{\rm R} = \rho_{\gamma} + \rho_{e} +
3\rho_{\nu} = {43 \over 8}\rho_{\gamma},
\label{rho0}
\ee
where $\rho_{\gamma}$ is the energy density in the CBR photons (which, 
today, have redshifted to become the cosmic background radiation photons 
observed at a temperature of 2.7K). In this case, the time-temperature 
relation (derived from the Friedman equation) is, 
\be
{\rm Pre-\epm annihilation}:~~t~T_{\gamma}^{2} = 0.738~{\rm MeV^{2}~s}.
\label{ttpre}
\ee

In SBBN it is often assumed that the neutrinos are fully decoupled prior 
to \epm annihilation and that they don't share in the energy transferred 
from the annihilating \epm pairs to the CBR photons.  In this approximation, 
the photons are hotter than the neutrinos in the post-\epm annihilation 
universe by a factor $T_{\gamma}/T_{\nu} = (11/4)^{1/3}$, and the 
relativistic energy density is
\be
\rho_{\rm R} = \rho_{\gamma} + 3\rho_{\nu} = 1.68\rho_{\gamma},
\ee
corresponding to a new time-temperature relation,
\be
{\rm Post-\epm annihilation}:~~t~T_{\gamma}^{2} = 1.32~{\rm MeV^{2}~s}.
\label{ttpost}
\ee

In quite general terms, one possible consequence of new physics beyond 
the standard model could be a non-standard, early Universe expansion 
rate, parameterized by an expansion rate factor $S$,
\be
H \rightarrow H' \equiv SH.
\label{S}
\ee
Although the introduction of $S$ here is quite general, in practice 
one must adopt a specific scheme through which this occurs.  Such a 
non-standard expansion rate might, but need not, be due to a change 
in the energy density: a change in the strength of gravity would also 
alter the expansion rate of the early Universe \cite{KS2003} as would 
non-standard, higher dimensional models which modify the expansion 
rate -- energy density relation (the Friedman equation, eq.~1) in our 
3+1 dimensional world \cite{rs}.  Different gravitational couplings 
for fermions and bosons \cite{BS2004} would have similar effects.  
These different mechanisms for implementing a non-standard expansion 
rate are not necessarily equivalent. 

If consideration is restricted to only the possibility of additional 
energy density, then 
\be
\rho_{\rm R} \rightarrow
\rho_{\rm R}' \equiv S^{2}\rho_{\rm R},
\label{rho'}
\ee
where $\rho_{\rm R}' = \rho_{\rm R} + \rho_{X}$ and $X$ identifies 
the extra component.  With the further restriction that the $X$ are 
also relativistic, this extra component behaves just like an additional 
neutrino though we emphasize that ``X" need not be additional flavors 
of active or sterile neutrinos.  In these circumstances S is a constant 
prior to \epm annihilation and it is convenient to account for the 
extra contribution to the standard-model energy density by normalizing 
it to that of an ``equivalent" neutrino flavor \cite{ssg}, so that 
\be
\rho_{X} \equiv \Delta N_{\nu}\rho_{\nu} =
{7 \over 8}\Delta N_{\nu}\rho_{\gamma}.
\label{deln}
\ee
For this case,
\be
S \equiv S_{pre} = (1 + {7 \over 43}\Delta N_{\nu})^{1/2}.
\label{sdeltannu}
\ee
However the expansion rate is more fundamental to BBN than is 
$\Delta N_{\nu}$, so we parameterize this class of non-standard 
models using $S$ (but, for comparison, we will often also quote 
the corresponding value of \Deln as given by eq.~\ref{sdeltannu}).

Not every additional energy density component can be accommodated 
this way since we are requiring that $\rho_{X}$ is proportional to 
$\rho_{\nu}$ as the Universe evolves.  An example that breaks this 
mold is if the ``X" were not `decoupled', in the sense that $X$ 
shared in the energy released when the \epm pairs annihilate.  
Other examples are found within some quintessence models 
\cite{SDTY1992,SDT1993,BHM2001,KS2003}.

As we discussed above, D, \3he, and \7li act as the principle baryometers 
for BBN since the \4he relic abundance does not vary considerably with 
$\eta$.  But \4he is very sensitive to the competition between the weak 
interaction rates (interconverting neutrons and protons) and the universal 
expansion rate which, during the early, radiation dominated evolution is 
fixed by the energy density in relativistic particles (``radiation'').  
As a result, while D, \3he, and \7li probe the baryon density, consistency 
between their abundances and that of \4he {\it tests} the standard model 
and provides {\it constraints} on (and, perhaps, hints of) physics beyond 
the standard model.  


\subsection{Lepton Asymmetry: $\xi$}

The baryon-to-photon ratio $\eta = n_{\rm B}/n_{\gamma}$ provides a 
dimensionless measure of the universal baryon asymmetry which is very 
small ($\eta ~\la 10^{-9}$).  By charge neutrality the asymmetry in 
the charged leptons must also be of this order.  However, there are no 
direct observational constraints (see \cite{ks92,kssw01,barger03b} and 
further references therein), on the magnitude of any asymmetry among the 
neutral leptons (neutrinos).  A dimensionless measure of the magnitude 
of the neutral lepton asymmetry is provided by $\xi$, the ratio of the 
neutral lepton chemical potential to the temperature (in energy units): 
$\xi \equiv \mu/kT$.  For example, for a neutrino flavor $\alpha$, the 
asymmetry (``neutrino degeneracy'') $L_\alpha$, between the numbers 
of $\nu_\alpha$ and $\bar\nu_{\alpha}$ is 
\be
L_\alpha\equiv {n_{\nu_{\alpha}}-n_{\bar\nu_{\alpha}} \over n_\gamma}=
{\pi^2 \over 12 \zeta(3)}\bigg(\xi_\alpha+{\xi_\alpha^3 \over \pi^2}\bigg)\,.
\ee
Note that for $\xi_\alpha \ll 1$, $L_\alpha \approx 0.684 \xi_\alpha$.
Mixing among the three active neutrinos ($\nu_{e}, \nu_{\mu}, \nu_{\tau}$)
ensures that at the time of BBN, $L_{e} \approx L_{\mu} \approx L_{\tau}$
($\xi_{e} \approx \xi_{\mu} \approx \xi_{\tau}$) \cite{equal}.

While any neutrino degeneracy ($\xi_{\alpha} < 0$ as well as $> 0$) 
will {\it increase} the energy density in the relativistic neutrinos 
and, hence, the early Universe expansion rate, the discussion here is 
limited to sufficiently small values of $\xi_{\alpha}$ so that the 
effect on $S$ of a non-zero $\xi_{\alpha}$ is negligible. Even so, a 
non-zero but relatively small asymmetry between {\it electron} type 
neutrinos and antineutrinos ($\xi_{e} ~\ga 10^{-2}$), while large 
compared to the baryon asymmetry, can still have a significant impact 
on the early Universe. The interconversion of neutrons to protons is 
affected by the $\nu_{e}$ so that any non-zero $\xi_{e}$ will alter 
the n/p ratio, thereby affecting the yields of the light nuclides 
formed during BBN.  

Of the light nuclei, the neutron limited \4he abundance is the most 
sensitive to $\xi_{e}$ and, together with the D, \3he, and \7li 
baryometers, again provides a test of the consistency of the 
standard model. 


\subsection{BBN Predictions}

\begin{figure}[htbp]
\begin{center}
\epsfxsize=3.4in
\epsfbox{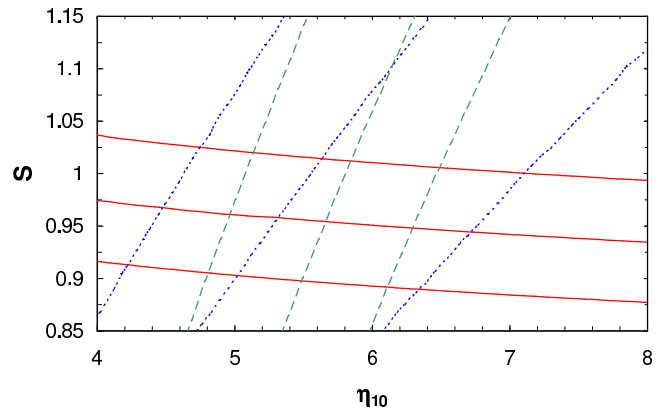}
      \caption{BBN-predicted isoabundance curves for D (dotted), \7li
      (dashed), and \4he (solid) in the expansion-rate parameter ($S$),
      baryon density parameter ($\eta$) plane.  From left to right, for
      D, \yd = 4, 3, 2, while for \7li, \yli = 3, 4, 5.  For \4he,
      from bottom to top, \Yp = 0.23, 0.24, 0.25.
      \label{svseta}}
\end{center}
\end{figure}

\begin{figure}[htbp]
\begin{center}
\epsfxsize=3.4in
\epsfbox{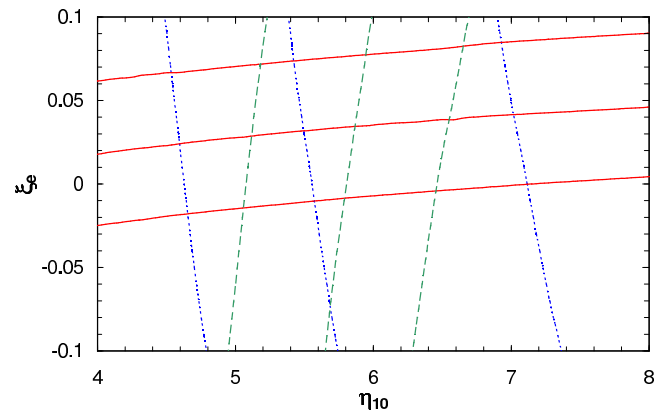}
      \caption{BBN-predicted isoabundance curves for D (dotted), \7li
      (dashed), and \4he (solid) in the lepton asymmetry parameter 
      ($\xi_{e}$), baryon density parameter ($\eta$) plane.  From 
      left to right, for D, \yd = 4, 3, 2, and for \7li, \yli 
      = 3, 4, 5.  For \4he, from top to bottom, \Yp = 0.23, 0.24, 0.25.
      \label{xivseta}}
\end{center}
\end{figure}

The BBN-predicted D, \4he, and \7li isoabundance curves for $S$ versus 
$\eta$ and for \xie versus $\eta$ are shown in Figures 1 and 2.  For 
D, the curves represent the ratio (to hydrogen) by number \yd $\equiv 
10^{5}$(D/H) = 2, 3, 4; similarly, for \7li, the curves correspond to 
\yli $\equiv 10^{10}$(Li/H) = 3, 4, 5; for \4he, the curves correspond 
to the mass fraction \Yp = 0.23, 0.24, 0.25.  These figures show clearly 
that D and \7li are good baryometers, sensitive to $\eta$, while \4he 
is a good chronometer, sensitive to $S$, as well as a good leptometer, 
sensitive to $\xi_{e}$.


\section{Observed Primordial Abundances}\label{sec:primabund}

The success of Big Bang Nucleosynthesis relies on its ability to predict 
the observed primordial abundances and, conversely, to learn about 
the cosmological parameters using these same abundances.  However 
the comparison of predictions and observations is far from trivial 
because of the further processing of the nuclei since the end of BBN.  

Deuterium is the baryometer of choice, primarily because since BBN 
its observed abundance should have only decreased as gas is cycled
through stars where D is burned to heavier, more tightly bound, 
nuclei \cite{els}.  As a result, D observed anywhere in the Universe,
at any time during its evolution, should provide a {\it lower} bound
to its primordial abundance.  Further, the deuterium observed in the 
high redshift, low metallicity QSO absorption line systems (QSOALS) 
should be very nearly primordial.  In contrast, the post-BBN evolution 
of \3he and of \7li are considerably more complicated, involving the 
competition between production, destruction, and survival.  As a result, 
at least so far, the current, locally observed (in the Galaxy) abundances 
of these nuclides have been of less value than has that of deuterium. 
Indeed, over the range in \yd to be adopted below, the primordial 
abundance of \3he is predicted to lie in the narrow range, $1.0 ~\la 
y_{3} ~\la 1.1$, in excellent agreement with that inferred from Galactic 
observations~\cite{rood}.  Thus, \3he provides similar, but less compelling 
constraints than does D.  While current estimates of primordial lithium, 
based on observations of very metal-poor (nearly primordial) stars, are 
problematic, \7li is retained here in order to highlight this challenge.  
For \4he the post-BBN evolution is straightforward, but difficult to 
quantify at a sufficient level of accuracy.  As gas is cycled through 
stars, hydrogen is burned to helium; only a very small fraction of the 
helium is burned to heavier nuclei.  If material is sampled ``here and 
now", a correction based on models of the chemical evolution of galaxies 
would need to be made for post-BBN produced \4he.  To avoid the model-dependent 
uncertainties associated with such corrections, which would surely 
overwhelm the observational errors, the best strategy is to search 
for \4he in the least evolved objects, in this case the metal-poor, 
extragalactic \hii regions.

Inferring the primordial D abundance from the QSOALS has not been without 
its difficulties, with some of the original abundance claims having been 
withdrawn or revised.  Presently there are 5 -- 6 QSOALS with reasonably 
firm deuterium detections \cite{bta,btb,o'm,pb,dod,k}.  However, there is 
significant dispersion among the abundances and the data fail to reveal 
the anticipated ``deuterium plateau" at low metallicity or at high redshift 
\cite{gs01}.  Here we adopt the five abundance determinations collected 
in the recent paper of Kirkman \etal \cite{k}.  The weighted mean value 
of \yd is 2.6~\footnote {This differs from the result quoted in Kirkman 
\etal because they have taken the mean of log($y_{\rm D}$) and then 
used it to infer \yd (\yd $ \equiv 10^{<log(y_{\rm D})>}$).}.  The 
dispersion among these five data points is very large, resulting in a 
$\chi^{2} = 15.3$ for four degrees of freedom, suggesting that one or 
more of these abundance determinations may be in error, or affected by 
unidentified and unaccounted for systematic errors.  To allow for this, 
we follow the approach advocated by \cite{o'm} and \cite{k} and adopt 
for the uncertainty in \yd the dispersion among the data points divided 
by the square root of the number of them.  Thus, the primordial abundance 
of deuterium to be used here is chosen to be: \yd $ = 2.6 \pm 0.4$.  

A somewhat less than clear situation also exists for the determinations 
of the primordial abundance of \4he.  There have been two, largely 
independent, estimates of \Yp based on analyses of large data sets of 
low-metallicity, extragalactic \hii regions.  The ``IT" \cite{itl,it} 
estimate of Y(IT) $= 0.244 \pm 0.002$, and the ``OS" determination 
\cite{os,oss,fo} of Y(OS) $= 0.234 \pm 0.003$ which differ by nearly 
$3\sigma$.  Very recently, IT have expanded their original data set 
and attempted to account for some (but not all) of the systematic 
uncertainties~\cite{it04}; see, \eg~\cite{os04}.  For their full data 
set of 89 \hii regions, IT find \Yp $= 0.2429\pm0.0009$.  To provide
a contrast with previous results, this value is adopted for our
analysis here. 

As with \4he, the abundance of \7li grows in the course of post-BBN
evolution, with lithium being produced in some stars (at some distinct
times in their evolution) as well as in cosmic ray spallation reactions
(where CNO nuclei are broken down to, among other nuclides, lithium).
So, as for D and \4he, lithium should be observed in the least evolved,
most metal-poor objects.  While lithium has been observed in the Sun
and in the solar system, as well as in the interstellar medium of the
Galaxy, these provide evolved, metal-enriched samples of little use in
estimating the primordial abundance of lithium.  The most relevant data
comes from studies of the very metal-poor stars in the halo of the Galaxy
or in Globular Clusters.  While this material should be very nearly
primordial, it must be kept in mind that these are observations of the
surfaces of the oldest stars in the Galaxy.  If, during the course of
their evolution, the surface material of these stars were mixed with
the lithium-depleted interiors, the currently observed surface lithium
abundances may not reflect the lithium abundances in these stars at 
their birth.  The most recent data from studies of halo stars is from
Ryan \etal\cite{ryan} who find \yli $= 1.23^{+0.34}_{-0.16}$. In contrast, and 
in some conflict with this result, Bonifacio \etal\cite{bonif} derive
from a sample of Globular Cluster stars, \yli $= 2.19^{+0.46}_{-0.38}$.

Below we compare each of these estimates of the primordial abundances with
the BBN predictions, standard as well as non-standard, fixed by the
adopted D and \4he abundances.


\section{Fits}

The non-linear, coupled differential equations of BBN are not conducive 
to analytic solution so that detailed comparison of the theoretical 
predictions with the observed/derived abundances of the light nuclei 
can only be achieved after resorting to a  lengthy process of numerical 
calculations. This necessity blurs the connection between the parameter
set \{$\eta$, $S$, $\xi_{e}$\} and the data set \{$y_{\rm D}$, Y$_{\rm P}$, 
$y_{\rm Li}$\}, and, conversely, the parameter values the data recommend. 
Yet it is clear from Figures 1  and 2 that the relic light nuclide abundances 
are smoothly varying, monotonic functions of $\eta$, $S$, and $\xi_{e}$, 
so it isn't surprising that over limited but substantial ranges, simple 
relations exist between the predicted abundances and these parameters. 
While the BBN-predicted primordial abundances are certainly {\it not} 
linearly related to the baryon density, the expansion rate, or the lepton 
asymmetry, our goal here is to find simple, linear fits to the predicted 
abundances (or powers of them) as functions of these parameters (to be 
defined more carefully below).  These fits work well, sometimes remarkably 
well, over limited ranges in the parameters (and/or over limited ranges 
in the predicted abundances of D, \4he, and \7li).

Since the domains over which our fits are applicable are restricted,
we must focus upon specific values for $\eta$, $S$ and $\xi_{e}$.  
The ``target" value/range of the baryon density is motivated by the 
(non-BBN) results from the WMAP constraints from the CBR temperature 
fluctuations \cite{sperg} where, for SBBN, $\eta_{10} = 6.14 \pm 0.25$.  
As shown in \cite{barger03a,barger03b} (and references therein), this 
estimate is little affected by a non-standard expansion rate and/or any 
(small) lepton asymmetry.  As a result, a limited range in $\eta_{10}$, 
centered around $\eta_{10} = 6$, is chosen here: $4 \leq \eta_{10} \leq 
8$. For our simple, linear fits to the BBN-predicted abundances as a 
function of $S$ to be sufficiently accurate, we must restrict the range 
in $S$.  This is the case for the fits we adopt provided that $0.85 
\leq S \leq 1.15$ corresponding to $-1.7 ~\la \Delta$N$_{\nu} ~\la 
2.0$ ($1.3 ~\la $N$_{\nu} ~\la 5.0$).  For an ``interesting" range in 
\xie over which the fits are reasonably accurate, $-0.1 \leq \xi_{e} 
\leq 0.1$ is adopted.

Before presenting our fits, it is worth reemphasizing that Figures 1 
\& 2 show clearly that D and \7li are good baryometers, sensitive to 
$\eta$, while \4he is a good chronometer, sensitive to $S$, as well 
as a good leptometer, sensitive to $\xi_{e}$.


\subsection{~Helium-4}

\begin{figure}[htbp]
\begin{center}
\epsfxsize=3.4in
\epsfbox{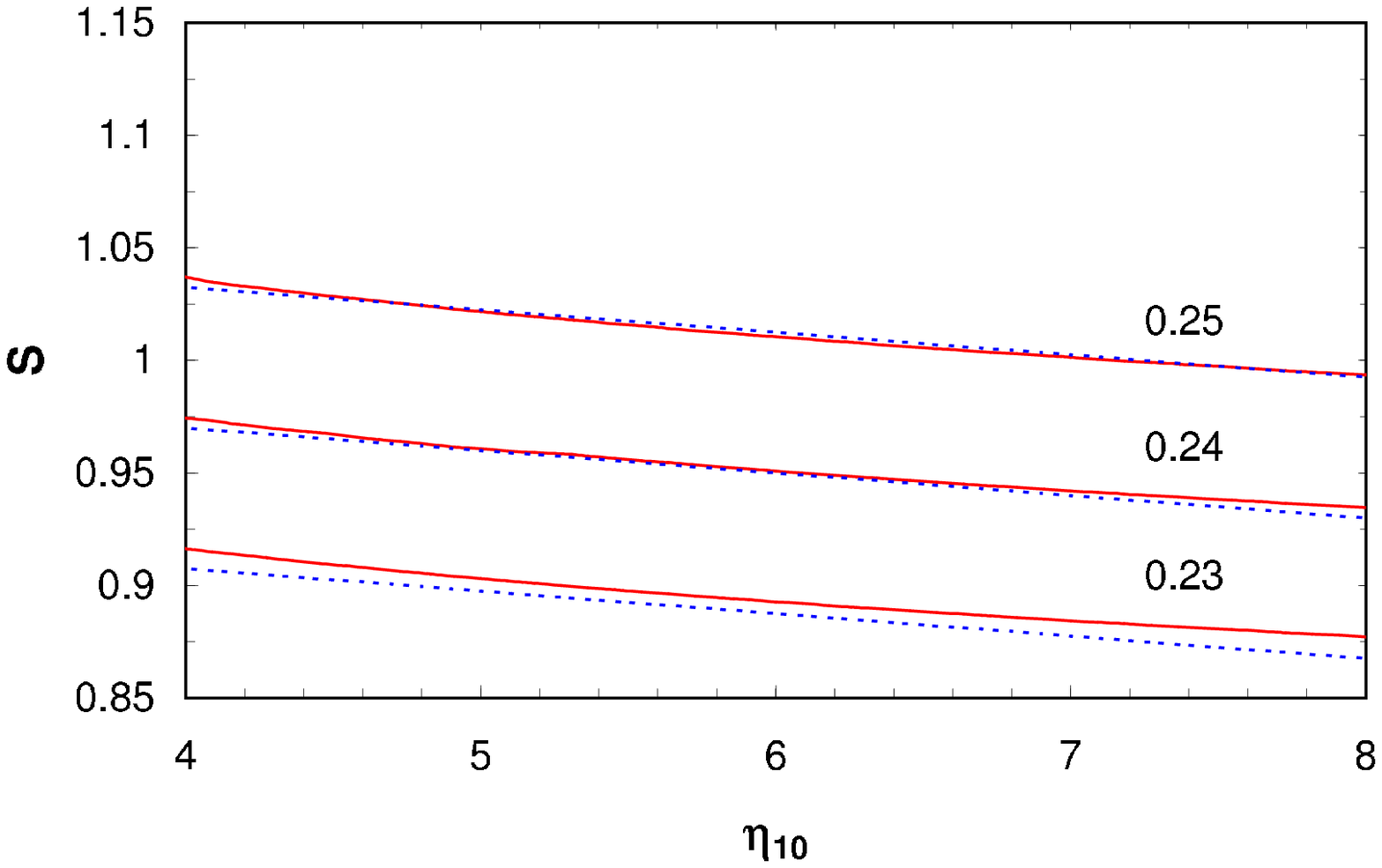}
      \caption{BBN-predicted isoabundance curves for \4he in the
      expansion-rate parameter ($S$), baryon density parameter 
      ($\eta$) plane.  From bottom to top, \Yp = 0.23, 0.24, 0.25.
      The solid curves are the BBN-predicted results, while the
      dotted curves are our fits (see the text).
      \label{svsetahe}}
\end{center}
\end{figure}
\begin{figure}[htbp]
\begin{center}
\epsfxsize=3.4in
\epsfbox{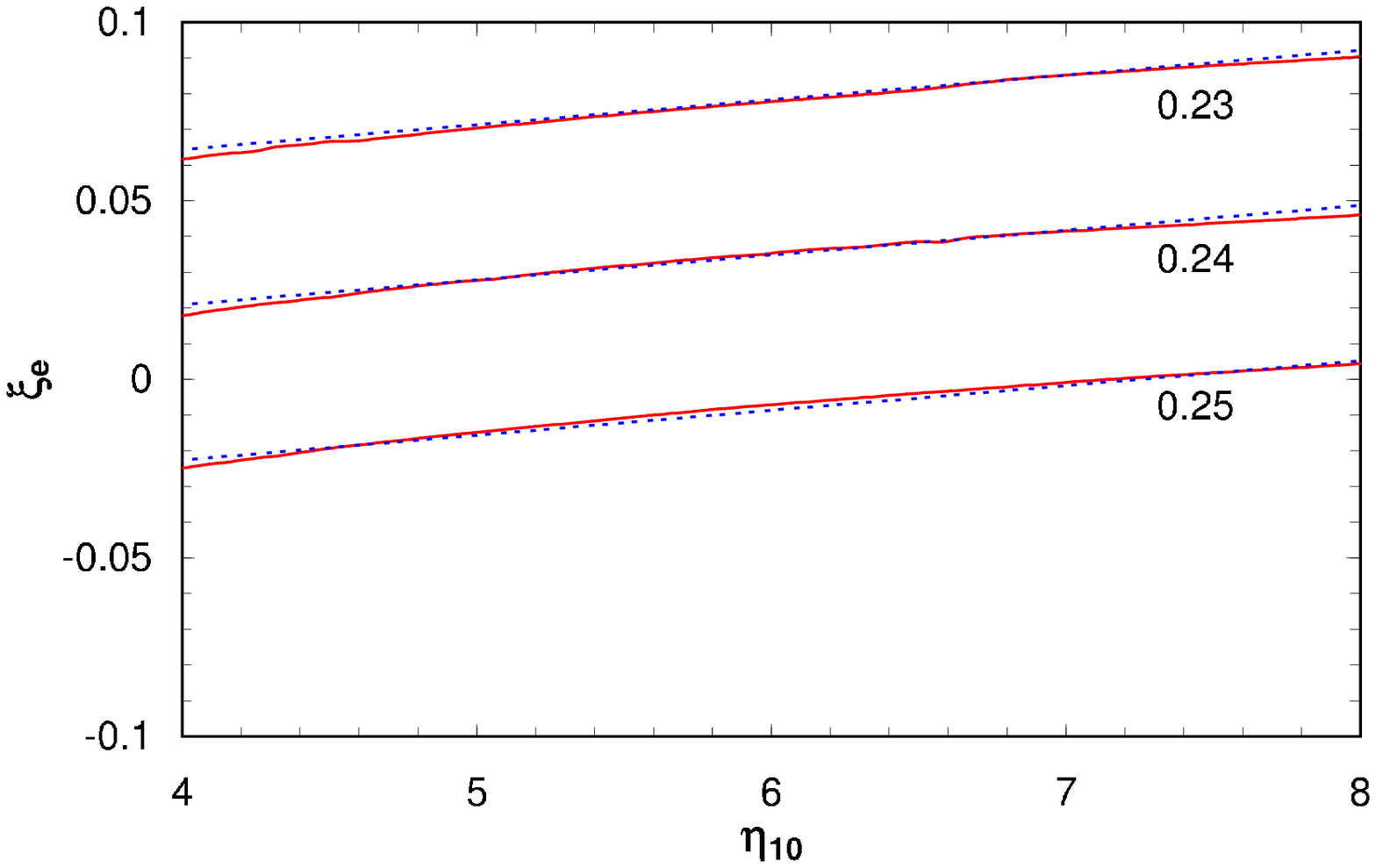}
      \caption{BBN-predicted isoabundance curves for \4he (solid) 
      in the lepton asymmetry parameter ($\xi_{e}$), baryon density 
      parameter ($\eta$) plane.  From top to bottom, \Yp = 0.23, 
      0.24, 0.25. The solid curves are the BBN-predicted results, 
      while the dotted curves are our fits (see the text).
      \label{xivsetahe}}
\end{center}
\end{figure}

The \4he abundance (mass fraction: Y$_{\rm P}$) is very insensitive to 
the baryon density at BBN, varying roughly logarithmically with $\eta$ 
over the range of $\eta$ adopted here ($4~\la \eta_{10}~\la 8$).  Thus, 
while for SBBN ($S = 1$ and \xie = 0) to a very good approximation Y = 
Y($\eta) \propto ~$ln $\eta$, we prefer to find a {\it linear} relation 
between \Yp and $\eta$.  Motivated by simplicity, the following linear 
fit for Y versus $\eta$ agrees with the SBBN-predicted predicted abundance 
to within 0.0006 ($\la 0.25\%$) over our adopted range in $\eta$.
\be
Y_{\rm P}^{FIT} \equiv 0.2384 + 0.0016\eta_{10} = 0.2384 + \eta_{10}/625.
\ee
We note that over the same range in $\eta$ this fit also agrees with 
the BNT \cite{bnt} predictions for \Yp to within 0.0002 ($\la 0.1\%$) 
or better.  While the accuracy of this fit is certainly not perfect, 
the difference is still smaller than the current errors in the 
observationally inferred primordial value of Y$_{\rm P}$.

As with the \Yp -- $\eta$ relation, the following linear fits to 
the \Yp -- $S$ and \xie relations work very well over the adopted 
parameter ranges (see Figures 1, 2, \ref{svsetahe} \& \ref{xivsetahe}).
\be 
Y_{\rm P}^{FIT} \equiv 0.2384\pm 0.0006 + 0.0016\eta_{10} + 
0.16(S-1) - 0.23\xi_{e}.
\label{eq:yfit}
\ee
Notice that for {\bf fixed $\eta_{10}$} and $\Delta$N$_{\nu} \ll 1$, 
$\Delta$Y $\approx 0.013\Delta$N$_{\nu}$.

It is convenient to rewrite this fit of Y$_{\rm P}(\eta,S,\xi_{e}$) 
(eq.~\ref{eq:yfit}) in a form which will facilitate comparison 
with comparable fits to the y$_{\rm D}(\eta,S,\xi_{e}$) and 
y$_{\rm Li}(\eta,S,\xi_{e}$) relations.  To this end, we introduce 
$\eta_{\rm He}$, defined as follows,
\be
\eta_{\rm He} \equiv 625({\rm Y}_{\rm P} - 0.2384\pm0.0006).
\ee
Then,
\be 
\eta_{\rm He} = \eta_{10} + 100(S-1) - {575\xi_{e} \over 4}.
\label{eq:etahe}
\ee
The meaning of $\eta_{\rm He}$ is clear; $\eta_{\rm He}$ is the 
SBBN ($S = 1$ and \xie = 0) value of $\eta_{10}$ corresponding 
to the adopted value of Y$_{\rm P}$.  Once \Yp is chosen, the 
resulting value of $\eta_{\rm He}$ provides a linear constraint 
on the combination of $\eta$, $S$, and \xie shown in eq.~\ref{eq:etahe}.  
As may be seen in Figures 3 \& 4, this fit works well (for $4 
\leq \eta_{10} \leq 8$) for $0.23~\la $Y$_{\rm P} ~\la 0.25$, 
corresponding to $-5~\la \eta_{\rm He}~\la 7$.

Adopting the IT~\cite{it04} value \Yp $= 0.2429\pm0.0009$ (see
\S~\ref{sec:primabund}), leads to $\eta_{\rm He} = 2.81\pm0.68$ 
($\Omega^{\rm He}_{\rm B}h^{2} = 0.0103\pm0.0025$).  Note that 
if, indeed, the very small IT error estimate is realistic, then 
helium is a competitive baryometer to deuterium!  The comparisons 
of this fit with the results from our BBN code are shown in 
Figures \ref{svsetahe} \& \ref{xivsetahe}.  As these figures and 
eq.~\ref{eq:etahe} show, $\eta_{\rm He}$, which is {\it linear} 
in Y$_{\rm P}$, is very sensitive to $S$ and $\xi_{e}$, but
relatively insensitive to $\eta$.


\subsection{~Deuterium}

\begin{figure}[htbp]
\begin{center}
\epsfxsize=3.4in
\epsfbox{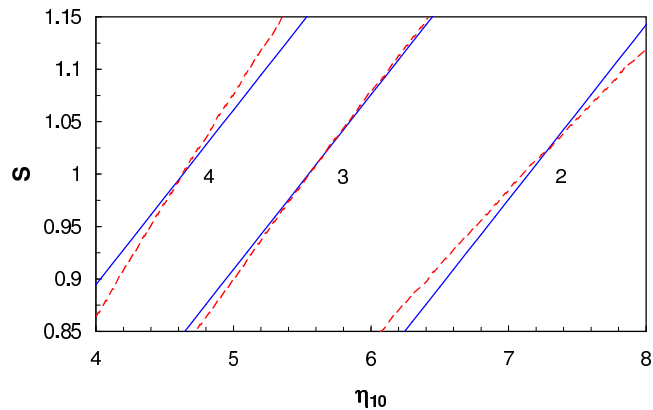}
      \caption{BBN-predicted isoabundance curves for D in the
      expansion-rate parameter ($S$), baryon density parameter 
      ($\eta$) plane.  From left to right, \yd = 4, 3, 2.
      The solid curves are the BBN-predicted results, while the
      dotted curves are our fits (see the text).
      \label{svsetad}}
\end{center}
\end{figure}
\begin{figure}[htbp]
\begin{center}
\epsfxsize=3.4in
\epsfbox{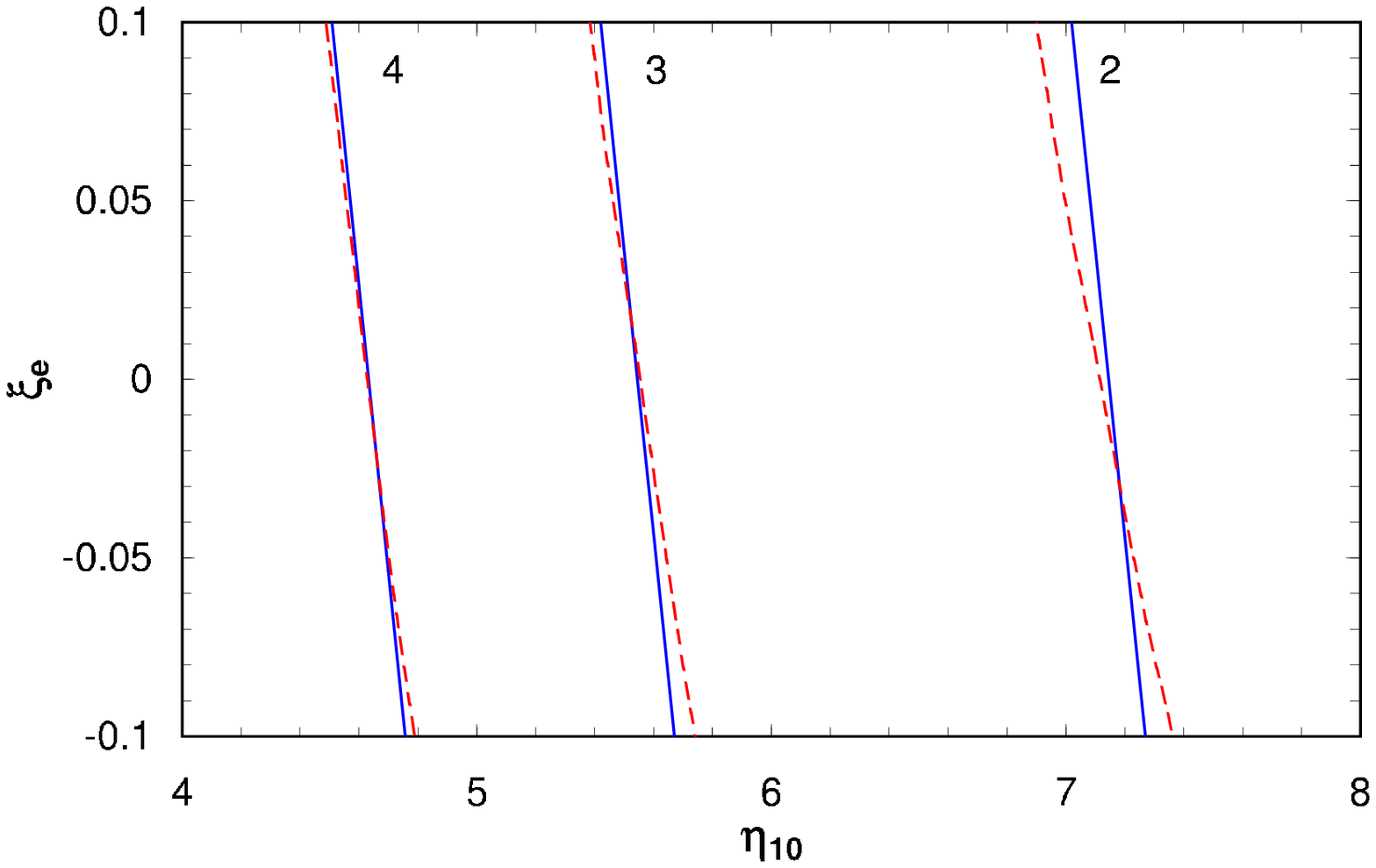}
      \caption{BBN-predicted isoabundance curves for D in the
      lepton asymmetry parameter ($\xi_{e}$), baryon density 
      parameter ($\eta$) plane.  From left to right, \yd = 4, 
      3, 2.  The solid curves are the BBN-predicted results, 
      while the dotted curves are our fits (see the text).
      \label{xivsetad}}
\end{center}
\end{figure}

Over the restricted range in $\eta$ under study here, the deuterium 
abundance \yd $\equiv 10^{5}($D/H), is well described by a power 
law in $\eta$ with \yd $\propto \eta^{-1.6}$.  For our adopted
range of $\eta$, \yd = $y_{\rm D}(\eta$) is well fit by
\be
y_{\rm D}^{FIT} \equiv 46.5\eta_{10}^{-1.6}.
\ee
While the true \yd -- $\eta$ relation is not precisely a power 
law, for $4 ~\la \eta_{10} ~\la 8$, this fit is accurate to better 
than 1\%, three times smaller than the $\sim 3\%$ BBN uncertainty 
estimated by BNT \cite{bnt}; this fit agrees with the BNT prediction 
to 2\% or better over this range in $\eta$.

In analogy with $\eta_{\rm He}$ defined by the \4he abundance 
(eq.~12), it is useful to define a similar parameter, $\eta_{\rm 
D}$, for D (including a $3\%$ error estimate),
\be
\eta_{\rm D} \equiv ({46.5(1\pm0.03) \over y_{\rm D}})^{1/1.6}.
\ee
Substituting the abundance estimate (\S\ref{sec:primabund}) for 
primordial D into this equation leads to our adopted value for 
$\eta_{\rm D} = 6.06^{+0.68}_{-0.53}$ ($\Omega^{\rm D}_{\rm B}h^{2} 
= 0.0221^{+0.0025}_{-0.0019}$). 

For the adopted ranges in $\eta$, $S$, and $\xi_{e}$, a simple 
linear relation which provides a good fit to $\eta_{\rm D}$ 
(see Figures \ref{svsetad} \& \ref{xivsetad}) is,
\be
\eta_{\rm D} = \eta_{10} - 6(S-1) + {5\xi_{e} \over 4}.
\label{eq:etad}
\ee
As may be seen from Figures 5 \& 6, this fit works quite well (for 
$4 \leq \eta_{10} \leq 8$) for $2~\la y_{\rm D}~\la 4$, corresponding
to $5~\la \eta_{\rm D} ~\la 7$.  Given the restricted ranges for $S$ 
and $\xi_{e}$, it is clear from Figures \ref{svsetad} \& \ref{xivsetad} 
and eq.~\ref{eq:etad} that D is a senstive baryometer ($\eta_{\rm D} 
\approx \eta_{10}$).


\subsection{Helium-3}

The dependence of the BBN-predicted abundance of \3he on the baryon
density, expansion rate, and lepton asymmetry is similar to that of 
D but, \3he is considerably less sensitive to them than is D.  Indeed, 
over the parameter ranges adopted here, the \3he abundance is well 
fit by $y_{3} \propto y_{\rm D}^{0.35}$.  As a result, along with its 
more complicated post-BBN evolution, \3he provides complementary but 
less compelling constraints than does D.  In our subsequent analysis, 
\3he is set aside in favor of D.


\subsection{Lithium-7}

\begin{figure}[htbp]
\begin{center}
\epsfxsize=3.4in
\epsfbox{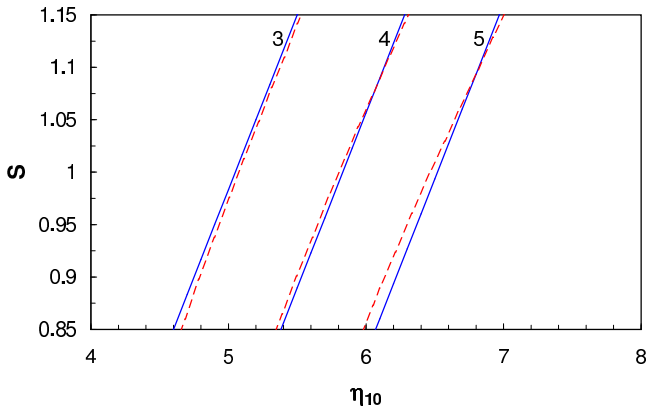}
      \caption{BBN-predicted isoabundance curves for \7li in the
      expansion-rate parameter ($S$), baryon density parameter 
      ($\eta$) plane.  From left to right, \yli = 3, 4, 5. The 
      solid curves are the BBN-predicted results, while the
      dotted curves are our fits (see the text).
      \label{svsetali}}
\end{center}
\end{figure}
\begin{figure}[htbp]
\begin{center}
\epsfxsize=3.4in
\epsfbox{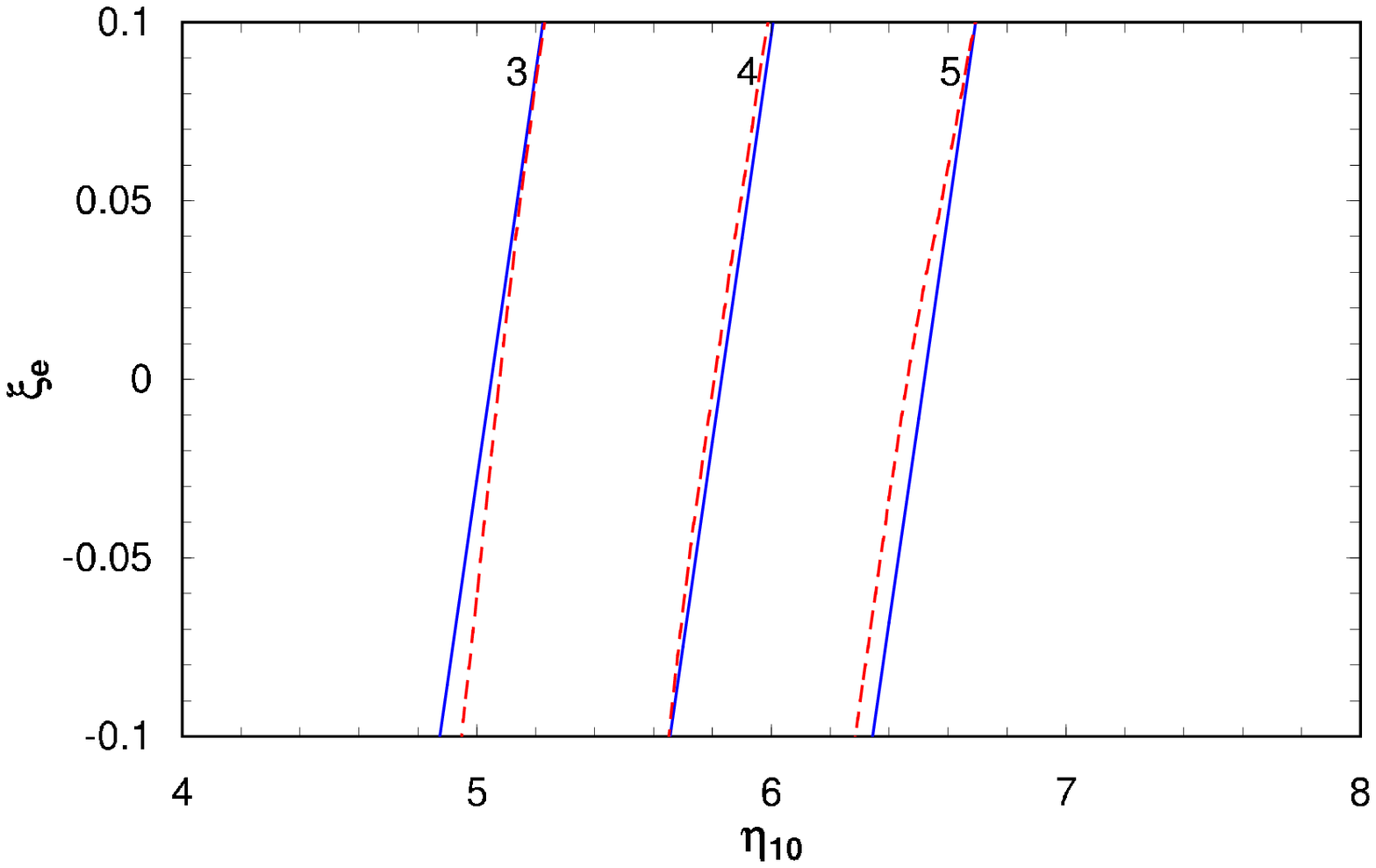}
      \caption{BBN-predicted isoabundance curves for \7li in the 
      lepton asymmetry parameter ($\xi_{e}$), baryon density parameter 
      ($\eta$) plane.  From left to right, \7li, \yli = 3, 4, 5. 
      The solid curves are the BBN-predicted results, while the 
      dotted curves are our fits (see the text).
      \label{xivsetali}}
\end{center}
\end{figure}

As with D, the \7li abundance\footnote{It is common in the astronomical 
literature to present the lithium abundance logarithmically: [Li]~$\equiv 
12 + $log(Li/H) = 2 + log(y$_{\rm Li}$).} is well described by a power 
law in $\eta$ over the range in baryon abundance explored here: \yli
$\equiv 10^{10}$(Li/H)$ ~\propto \eta^{2}$.  The following fit agrees 
with the BBN predictions to better than 3\% over the adopted range in 
$\eta$,
\be
y_{\rm Li}^{FIT} \equiv {\eta_{10}^{2} \over 8.5}.
\ee
While this fit predicts slightly smaller lithium abundances compared 
to those of BNT \cite{bnt}, the differences are at the 5-8\% level,
small compared to the BNT uncertainty estimates as well as those of 
Hata \etal~\cite{hata} ($\sim 10 - 20\%$).

In analogy with $\eta_{\rm D}$ and $\eta_{\rm He}$ defined above,
we introduce $\eta_{\rm Li}$ (allowing for a $10\%$ uncertainty), 
defined by,
\be
\eta_{\rm Li} \equiv (8.5(1\pm0.1)y_{\rm Li})^{1/2}.
\ee
Using the Ryan \etal\cite{ryan} estimate (\S\ref{sec:primabund}) 
\yli $= 1.23^{+0.34}_{-0.16}$, $\eta_{\rm Li} = 3.23^{+0.44}_{-0.28}$ 
($\Omega^{\rm Li}_{\rm B}h^{2} = 0.0118^{+0.0016}_{-0.0010}$), while 
the Bonifacio \etal\cite{bonif} result of \yli $= 2.19^{+0.46}_{-0.38}$ 
gives $\eta_{\rm Li} = 4.31^{+0.48}_{-0.46}$ ($\Omega^{\rm Li}_{\rm B}h^{2} 
= 0.0157\pm0.0017$).  

A simple, linear relation for $\eta_{\rm Li}$ as a function 
of $\eta$, $S$, $\xi_{e}$, which works reasonably well over 
the adopted parameter ranges (see Figures \ref{svsetali} \& 
\ref{xivsetali}) is,
\be
\eta_{\rm Li} = \eta_{10} - 3(S-1) - {7\xi_{e} \over 4}.
\label{eq:etali}
\ee
As Figures 7 \& 8 reveal, this fit works well (for $4 \leq \eta_{10} 
\leq 8$) for $3~\la y_{\rm Li}~\la 5$, corresponding to $5~\la 
\eta_{\rm Li}~\la 7$.  We note that this fit breaks down for \yli 
$\la 2$ ($\eta_{\rm Li} ~\la 4$).  In addition, Figures \ref{svsetali} 
\& \ref{xivsetali}, along with eq.~\ref{eq:etali}, show that, as 
is the case for deuterium, lithium can be an excellent baryometer 
(for the restricted ranges of $S$ and $\xi_{e}$, $\eta_{\rm Li} 
\approx \eta_{10}$).

  
\section{Applications}

In this section the application of our simple, linear fits (among 
$\eta$, $S$, $\xi_{e}$, and y$_{\rm D}$, y$_{\rm Li}$, Y$_{\rm P}$) 
is illustrated by considering several standard and nonstandard BBN 
options.

\subsection{SBBN} 
\begin{figure}[htbp]
\begin{center}
\epsfxsize=3.4in
\epsfbox{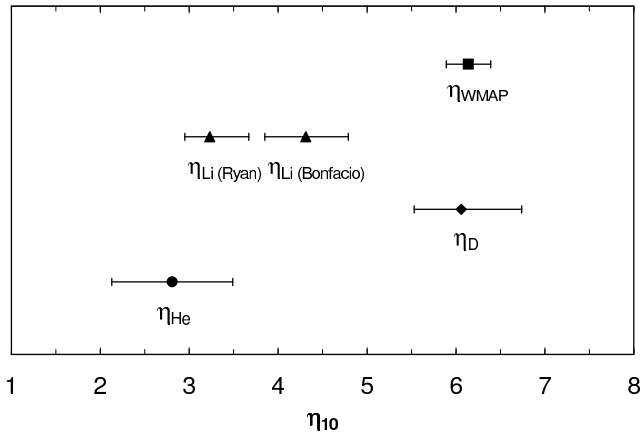}
      \caption{The baryon abundance parameters (see \S III) corresponding 
      to the D, \4he, and \7li abundances adopted here (from \S II), and 
      from WMAP~\cite{sperg}
      \label{etas}}
\end{center}
\end{figure}

As a first application of our simple fits to the predicted primordial
abundances, consider the case of SBBN ($S = 1$, $\xi_{e} = 0$).  
In this case the predicted abundances depend on only one adjustable
parameter, the baryon abundance parameter $\eta$.  For SBBN we may 
adopt the WMAP-inspired value \cite{sperg} of $\eta_{10} = 6.14\pm0.25$ 
(\obh $= 0.0224\pm0.009$).  For $S = 1$ and \xie = 0 we expect,
\be
SBBN:~\eta_{\rm D} = \eta_{\rm He} = \eta_{\rm Li} = \eta_{10}.
\label{eq:sbbn}
\ee
It is clear from Figure~\ref{etas}, using the abundances identified 
in \S II, that eq.~\ref{eq:sbbn} is {\bf not} satisfied ($\eta_{\rm D} 
\approx \eta_{10} \neq \eta_{\rm He} \approx \eta_{\rm Li}$).  However, 
SBBN {\bf does} work for deuterium: $\eta_{\rm D} = 6.06^{+0.68}_{-0.53} 
\approx \eta_{10} = 6.14\pm0.25$.  In addition, we note that for the 
WMAP baryon abundance the SBBN-predicted value of the \3he primordial 
abundance, $1.1\pm0.1\times 10^{-5}$, is in excellent agreement with 
that inferred from the study of Galactic \hii regions \cite{rood}.  
The problems arise for \4he and \7li.

For \4he, the recent Izotov \& Thuan~\cite{it04} determination of 
the primordial mass fraction leads to $\eta_{\rm He} = 2.81\pm0.68$
(\S IIIA), which is $\sim 6\sigma$ below the WMAP baryon density 
parameter.  While Izotov \& Thuan have attempted to account for some 
of the potential systematic errors in their \Yp determination, it 
may well be that they have underestimated the residual error; see, 
\eg Olive and Skillman~\cite{os04}.  If not, this disagreement may 
provide a hint of new physics beyond the standard model ($S \neq 1$ 
and/or $\xi_{e} \neq 0$).

As may be seen from Figures 1, 2, 4, \& 7, along with eqs.~16 \& 19, 
the BBN-predicted abundance of \7li is closely tied to that of D,
y$_{\rm Li} \approx 4.3(y_{\rm D}/2.6)^{-5/4}$.  For y$_{\rm D}$
in its observed range (and/or for $\eta_{10}$ in the WMAP range),
y$_{\rm Li} \approx 4 - 5$ ([Li] ~$\approx 2.6 - 2.7$).  For the 
WMAP baryon abundance the SBBN prediction is y$_{\rm Li} = 
4.44^{+0.58}_{-0.57}$, to be compared with the halo star estimate 
from Ryan \etal~\cite{ryan} of y$_{\rm Li} = 1.23^{+0.34}_{-0.16}$, 
or with the globular cluster estimate from Bonifacio \etal~\cite{bonif} 
of y$_{\rm Li} = 2.19^{+0.46}_{-0.38}$.  The SBBN-predicted \7li abundance 
is much higher, by factors of $\sim 2 - 4$, than those inferred 
from the studies of the oldest halo and/or globular cluster stars 
in the Galaxy.  It will be seen in more detail below that it is 
unlikely that any of the ``simple" nonstandard effects ($S \neq 1$, 
$\xi_{e} \neq 0$) can resolve this conflict (since the \7li and D 
abundances are so tightly coupled).  It may well be that for these 
oldest stars in the Galaxy, mixing surface material with the 
lithium-depleted interior has reduced their observed, surface 
lithium abundances from the primordial value~\cite{pwsn,pswn}.

\subsection{Non-Standard Expansion: $S \neq 1$, $\xi_{e} = 0$}

The previous section has reminded us of the tension between the
observationally-inferred primordial abundances of D, \4he, and 
\7li; see, \eg \cite{osw}.  While it is not unlikely that the 
resolution of these conflicts may be found in systematic uncertainties 
of the astronomy (\eg corrections to the derived helium abundance 
and/or lithium depletion/dilution via mixing), they may be providing 
hints of new physics.  Since the BBN abundance of \4he is sensitive 
to the early-Universe expansion rate and that of D probes the baryon 
density, a combination of the two abundances permits us to investigate 
constraints on models with non-standard expansion rates ~\cite{ssg,ossw}.  
Setting aside \7li for the moment and only using D and \4he, we may use 
equations 13 and 16 to find
\be
S - 1 = {\eta_{\rm He} - \eta_{\rm D} \over 106}
\ee
and,
\be
\eta_{10} = {100\eta_{\rm D} + 6\eta_{\rm He} \over 106}.
\ee
It is clear from eq.~23 that D is the dominant baryometer.  While it 
may appear from eq.~22 that D and \4he make comparable contributions
to constraining the expansion rate, recall that $\eta_{\rm He}$ is
{\bf linear} in \Yp while $\eta_{\rm D}$ varies only as a fractional
power of y$_{\rm D}$.  Thus, whereas reconciling \Yp with y$_{\rm D}$
requires an increase in \Yp of $\Delta$Y$_{\rm P} \approx 0.006$ 
($\sim 2.5\%$), reconciling D with \4he would require that y$_{\rm D}$
increase by a factor of $\sim 4 - 5$ (and, the resulting $\eta_{\rm D}$ 
would then be in conflict with the WMAP determination of $\eta_{10}$).  
From eq.~22, for $\Delta$N$_{\nu} \ll 1$ and with {\bf y$_{\rm D}$ 
($\eta_{\rm D}$) fixed}, $\Delta$Y$~\approx 0.014\Delta$N$_{\nu}$.  
Note that depending on the quantity held fixed ($\eta_{10}$ or 
$\eta_{\rm D}$), the coefficients in the $\Delta$Y -- $\Delta$N$_{\nu}$ 
relations differ slightly.

For the values adopted in \S IIIA,B (and their uncertainties),
$\eta_{\rm D} = 6.06^{+0.68}_{-0.53}$ and $\eta_{\rm He} = 2.81\pm 0.68$,
we find
\be
\eta_{10} = 5.88^{+0.64}_{-0.50} \ \ ; \ \ S = 0.969^{+0.008}_{-0.009}.
\ee

\begin{figure}[htbp]
\begin{center}
\epsfxsize=3.4in
\epsfbox{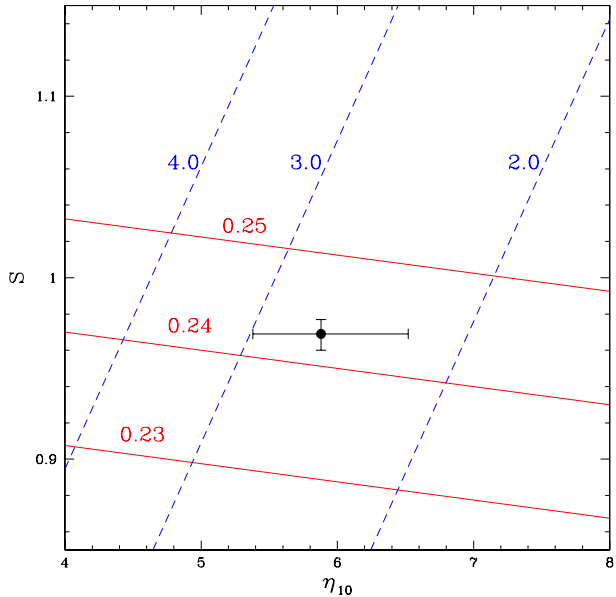}
      \caption{BBN-predicted isoabundance curves for D (dashed) and 
      \4he (solid) in the expansion-rate parameter ($S$), baryon 
      density parameter ($\eta$) plane.  The point with error bars
      corresponds to the adopted abundances for D and \4he ($\xi_{e}
      = 0$); see the text, eq.~23.
      \label{svseta2}}
\end{center}
\end{figure}
\begin{figure}[htbp]
\begin{center}
\epsfxsize=3.4in
\epsfbox{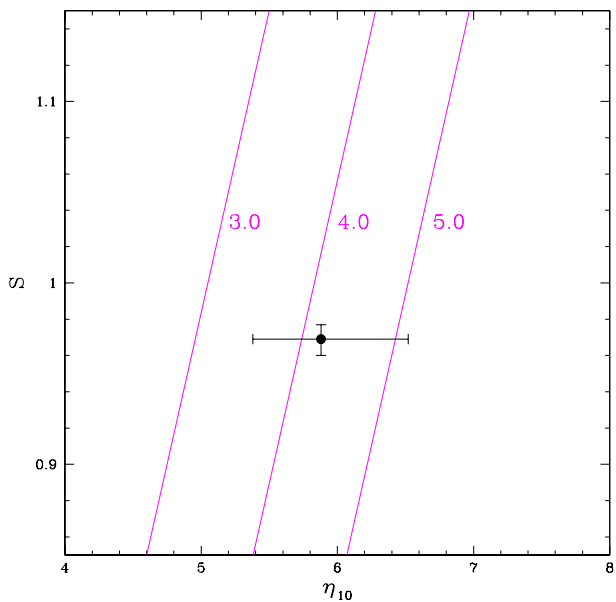}
      \caption{BBN-predicted isoabundance curves \7li in the 
      expansion-rate parameter ($S$), baryon density parameter 
      ($\eta$) plane.  The numbers next to the curves are for
      y$_{\rm Li}$.  The point with error bars corresponds to 
      the adopted abundances for D and \4he ($\xi_{e} = 0$) as
      in Figure 9.
      \label{svseta3}}
\end{center}
\end{figure}

If the non-standard expansion rate factor is associated with an 
equivalent number of neutrinos, $\Delta$N$_{\nu} = -0.37^{+0.10}_{-0.11}$.  
These results, along with the D and \4he isoabundance curves, are shown 
in Figure 10.  Notice that although \Deln is now closer to zero than in, 
\eg~\cite{barger03a}, the smaller adopted uncertainty in \Yp results in 
a larger, $\sim 3 - 4 \sigma$ discrepancy.  However, these BBN results 
for the baryon density and the expansion rate are in excellent agreement
with those from WMAP which, while sensitive to the baryon density
parameter is relatively insensitive to the expansion rate factor;
see~\cite{barger03a}.

Although the above combination of $\eta_{10}$ and $S$ can reconcile
the inferred primordial abundances of D and \4he, it cannot resolve
the lithium problem.
\be
\eta_{\rm Li} = {103\eta_{\rm D} + 3\eta_{\rm He} \over 106} = 
5.97^{+0.66}_{-0.52}.
\ee
Accounting for the theoretical uncertainty in the BBN lithium
production and in our fit,
\be
{\rm y}_{\rm Li} = 4.20^{+1.07}_{-0.81} \ \ ; \ \ [{\rm Li}]_{\rm P} 
= 2.62^{+0.10}_{-0.09}.
\ee
This conflict is shown in Figure 11 where the D and \4he constraints
on $\eta_{10}$ and $S$ are shown along with isoabundance curves for
\7li.  It is clear that y$_{\rm Li} < 3$ is strongly disfavored.

\subsection{~Non-Zero Lepton Number: $\xi_{e} \neq 0$, $S = 1$}

In an attempt to relieve the tension between the BBN predicted 
and observed deuterium and helium-4 abundances, a non-zero lepton 
number ($\nu_{e} - \bar{\nu}_{e}$ asymmetry; $\xi_{e} \neq 0$) is 
an alternative to the non-standard expansion rate explored in the 
previous section.  An excess of $\nu_{e}$ over $\bar{\nu}_{e}$ will 
drive down the pre-BBN neutron-to-proton ratio, leaving fewer neutrons 
to be incorporated in \4he.  The resulting, lower, BBN-predicted 
\4he abundance will be closer to that observed.  At the same time, 
the effect of an e-neutrino asymmetry on the abundances of D and/or 
\7li is small.
 
As for the case of $S \neq 1$, we set aside \7li and use D and \4he 
to constrain $\eta_{10}$ and $\xi_{e}$.  For $S = 1$,
\be
\eta_{10} = {115\eta_{\rm D} + \eta_{\rm He} \over 116}
\ee
and,
\be
\xi_{e} =  {\eta_{\rm D} - \eta_{\rm He} \over 145}. 
\ee

Using the D and \4he abundances adopted here,
\be
\eta_{10} = 6.04^{+0.67}_{-0.53} \ \  ; \ \  \xi_{e} = 0.022^{+0.007}_{-0.006}
\ee
\begin{figure}[htbp]
\begin{center}
\epsfxsize=3.4in
\epsfbox{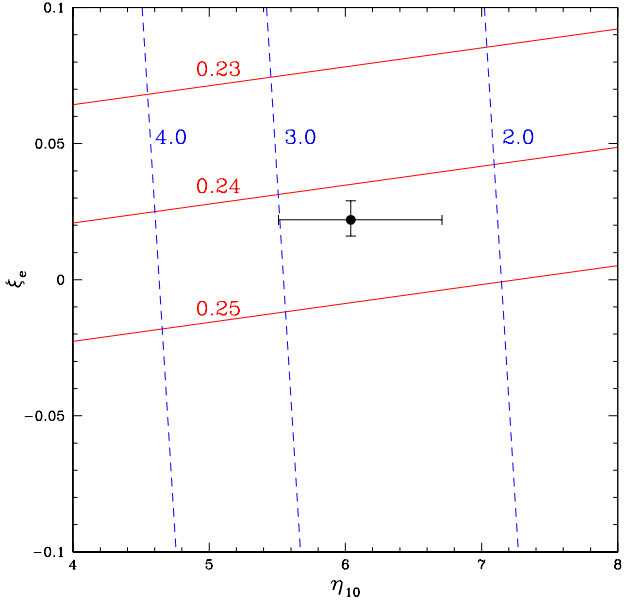}
      \caption{BBN-predicted isoabundance curves for D (dashed) and 
      \4he (solid) in the lepton asymmetry parameter ($\xi_{e}$), baryon 
      density parameter ($\eta$) plane.  The point with error bars
      corresponds to the adopted abundances for D and \4he ($S = 1$); 
      see the text, eq.~28.
      \label{xivseta2}}
\end{center}
\end{figure}

\begin{figure}[htbp]
\begin{center}
\epsfxsize=3.4in
\epsfbox{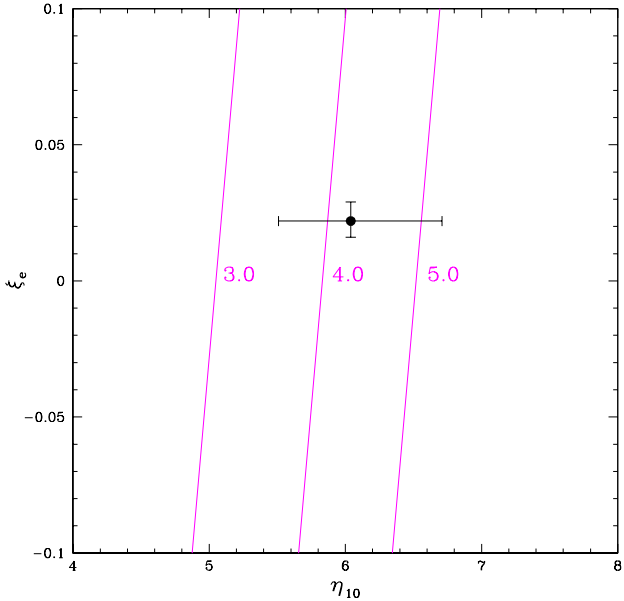}
      \caption{BBN-predicted isoabundance curves \7li in the 
      lepton asymmetry parameter ($\xi_{e}$), baryon density 
      parameter ($\eta$) plane.  The numbers next to the curves 
      are for y$_{\rm Li}$.  The point with error bars corresponds 
      to the adopted abundances for D and \4he ($S = 1$) 
      as in Figure 11.
      \label{xivseta3}}
\end{center}
\end{figure}

These results, along with the D and \4he isoabundance curves, are shown
in Figure 12.

A ``small" lepton asymmetry, $\xi_{e} \approx 0.02$, which, however, is 
very large compared to the baryon asymmetry ($\sim \eta~\la 10^{-9}$), 
can reconcile the observationally inferred primordial D and \4he abundances 
with the predictions of BBN.  But, this lepton asymmetry will {\bf not}
resolve the conflict between the BBN-predicted and observed \7li 
abundances since
\be
\eta_{\rm Li} = {142\eta_{\rm D} + 3\eta_{\rm He} \over 145} = 
5.99^{+0.67}_{-0.52}
\ee
and,
\be
{\rm y}_{\rm Li} = 4.23^{+1.08}_{-0.82} \ \ ; \ \ [{\rm Li}]_{\rm P} 
= 2.63^{+0.10}_{-0.09}.
\ee
As is the case for a non-standard expansion rate, the BBN-predicted
abundance of \7li is very tightly tied to that of D, and a non-zero
lepton number cannot resolve the conflict between theory and data.
This is illustrated in Figure 13, where it is clear that here, too, 
y$_{\rm Li} < 3$ is strongly disfavored.

\subsection{Non-Zero Lepton Number ($\xi_{e} \neq 0$) And Non-Standard 
Expansion Rate ($S \neq 1$; N$_{\nu} \neq 3$)}

In \S IVB it was seen that, in the absence of a non-zero lepton number, 
the D and \4he abundances severely restrict deviations of the early 
universe expansion rate from its standard value.  When associated with 
the effective number of neutrinos, this constraint (for the choice of 
D and \4he abundances adopted here) is nearly $4\sigma$ away from the 
standard value of N$_{\nu} = 3$.  If the LSND result \cite{lsnd} is 
interpreted in terms of a 4th, ``sterile", neutrino, the mixing of such 
a neutrino with the active neutrinos would bring it into equilibrium 
in the early universe prior to BBN~\cite{volkas}, resulting in N$_{\nu}
= 4$ at BBN.  Clearly (see \S IIIB), the current best estimates of
the primordial abundances cannot tolerate the corresponding speed-up
in the expansion rate at BBN.  However, in \S IIIC it was shown that
N$_{\nu} = 3$ is perfectly acceptable {\bf provided that} there is
a small but significant $\nu_{e} - \bar\nu_{e}$ asymmetry.  It is,
therefore, not unexpected that for $\xi_{e} \neq 0$, the BBN constraints
on $S$ ($\Delta$N$_{\nu}$) are considerably expanded (see, \eg Barger
\etal \cite{barger03a} and references therein).  In implementing 
for BBN $S \neq 1$ as well as $\xi_{e} \neq 0$, there is a problem.
Because of concerns about the primordial \7li (and \3he) abundances
as inferred from current observational data, the analysis here has 
been restricted to employing only the D and \4he abundances.  But,
with three free parameters ($\eta_{10}$, $S$ (or $\Delta$N$_{\nu}$), 
and $\xi_{e}$) and only two constraints ($\eta_{\rm D}$ and 
$\eta_{\rm He}$), non-BBN data (restricting $\eta_{10}$) is required
to proceed further.  Retaining the constraints from D and \4he, $S$
and \xie are then functions of the baryon density parameter.
\be
590(S-1) = 116\eta_{10} - (115\eta_{\rm D} + \eta_{\rm He})
\ee
and,
\be
145\xi_{e} = 106(S-1) + \eta_{\rm D} - \eta_{\rm He}.
\ee
For the abundances adopted here (\S\ref{sec:primabund}), this leads to
\be
590(S-1) = 116\eta_{10} - 700^{+78}_{-61},
\ee
and
\be
\xi_{e} = 0.731(S-1) + 0.0224^{+0.0066}_{-0.0059}.
\ee
This \xie -- $S$ relation is shown in Figure 14.

\begin{figure}[htbp]
\begin{center}
\epsfxsize=3.4in
\epsfbox{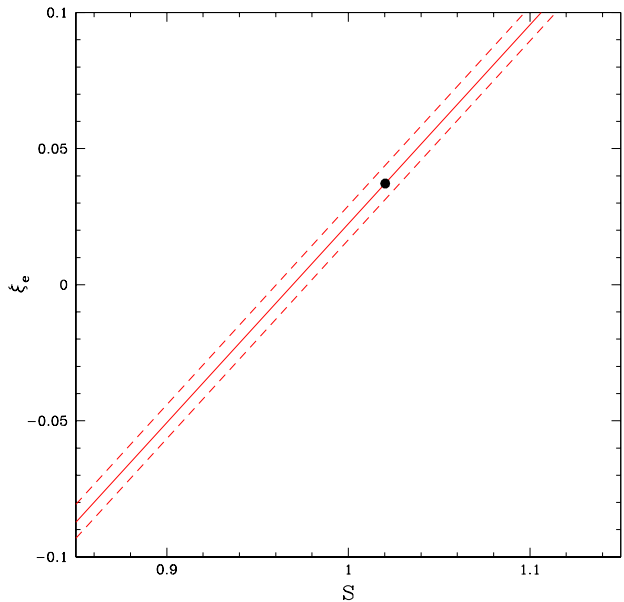}
      \caption{BBN-predicted $\xi_{e} - S$ relation corresponding to
      the adopted D and \4he abundances.  The solid curve is for the
      central value and the dashed curves represent the $1\sigma$ 
      uncertainties.  The point indicated corresponds to the WMAP
      value for the baryon density; see the text.
      \label{xivss}}
\end{center}
\end{figure}

A constraint on the baryon density is possible utilizing CBR data 
provided that the corresponding values of $S - 1$ and $\xi_{e}$ are 
sufficiently small, so that the expansion rate at recombination, some 
400 kyr after BBN, doesn't deviate significantly from the standard value.  
For the WMAP recommended baryon abundance~\cite{sperg}, $\eta_{10} = 
6.14$, $S \approx 1.0203$, corresponding to $\Delta$N$_{\nu} \approx 
0.25$ and $\xi_{e} \approx 0.037$.  This point is shown in Figure 14.

\begin{figure}[htbp]
\begin{center}
\epsfxsize=3.4in
\epsfbox{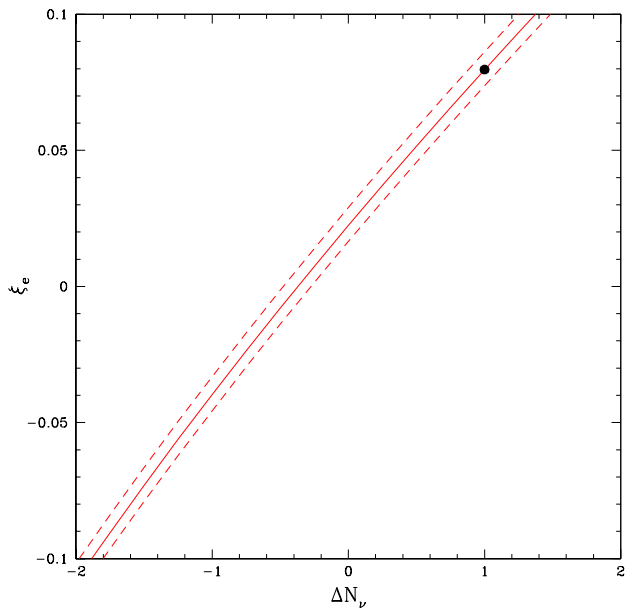}
      \caption{BBN-predicted $\xi_{e} - \Delta$N$_{\nu}$ relation 
      corresponding to the adopted D and \4he abundances.  The 
      solid curve is for the central value and the dashed curves 
      represent the $1\sigma$ uncertainties.  The point indicated 
      corresponds to N$_{\nu} = 4$.
      \label{xivsnnu4}}
\end{center}
\end{figure}

It is now possible to address the question of whether or 
not BBN permits N$_{\nu} = 4$ ($S = 1.0783$).  This choice 
for $S$ requires $\xi_{e} = 0.080$ and $\eta_{10} = 6.43$, 
only $\sim 1\sigma$ away from the WMAP value.  This point 
is shown in Figure 15 where the $\xi_{e} - \Delta$N$_{\nu}$ 
relation is displayed.  While BBN constraints on more extreme 
cases (\eg $\Delta$N$_{\nu} ~\ga 2$) can be explored (see, 
\eg Barger \etal~\cite{barger03b}), our simple fits begin 
to become inaccurate for $\xi_{e} ~\ga 0.1$, limiting our 
analysis here to $\Delta$N$_{\nu} ~\la 1.2$ (see Figure 15).

For the general case studied here, the BBN-predicted lithium
abundance depends on the baryon density (as well as on the
D and \4he abundances).
\be
\eta_{\rm Li} = 0.159\eta_{10} + 0.822\eta_{\rm D} + 
0.019\eta_{\rm He}.
\ee
Notice that here, too, the \7li abundance is tightly tied to 
that of D.  Indeed,
\be
\eta_{\rm Li} = \eta_{\rm D} + 3[(S-1) - \xi_{e}] \approx \eta_{\rm D}.
\ee  
As before, there are lithium ``problems" for the two cases studied 
above.  For the WMAP combination of parameters (see eq.~19), 
$\eta_{\rm Li} = 6.01$, so that \yli = 4.25 ([Li] = 2.63).  For 
N$_{\nu} = 4$, $\eta_{\rm Li} = 6.06$, corresponding to \yli = 
4.32 ([Li] = 2.64).  At least for these combinations of non-standard 
physics (as well as for SBBN), there is an apparent conflict between 
the BBN predicted and observed primordial abundances of lithium.


\section{Summary And Conclusions}

In the toolbox of a particle phenomenologist, BBN is a venerable and 
valuable tool, providing complementary constraints on -- or hints 
of -- new physics beyond the standard model.  As the models change
and the data are updated, not every phenomenologist on the street
may have easy access to a BBN code to enable quick revision of previous 
bounds or investigation of new hints.  It may, then, be of value to 
have simple fits to the BBN-predicted abundances as functions of 
the key variables (baryon density, expansion rate or neutrino number,
neutral lepton asymmetry).  For interesting but limited ranges of 
the parameters such fits have been presented here and applied to 
several simple examples.  We have seen that the observed D and 
\4he abundances, while apparently inconsistent with SBBN (modulo
systematic errors in the data), can be reconciled if non-standard
expansion rates and/or non-zero lepton number are allowed.  It was 
also shown how non-zero lepton number allows one -- or more -- sterile
neutrinos while maintaining the consistency of BBN.  Finally, by
revealing the clear connections between the BBN-predicted abundances
of D and \7li, we showed that none of these non-standard physics
solutions can reconcile the observed and predicted lithium abundances.


\end{document}